\documentclass[]{interact}

\usepackage{amsmath}
\usepackage{float}
\usepackage{epstopdf}
\usepackage[caption=false]{subfig}
\usepackage{listings}
\usepackage{etoolbox}

\usepackage{hyperref}

\usepackage[numbers,sort&compress,merge]{natbib}
\bibpunct[, ]{[}{]}{,}{n}{,}{,}

\theoremstyle{plain}

\theoremstyle{definition}

\theoremstyle{remark}

\begin{document}


\title{Towards Molecular Simulations that are Transparent, Reproducible, 
Usable By Others, and Extensible (TRUE)}


\author{
\name{Matthew W. Thompson, Justin B. Gilmer, Ray A. Matsumoto, Co D. Quach, Parashara Shamaprasad, Alexander H. Yang, Christopher R. Iacovella, Clare M$^{\rm c}$Cabe, and Peter T. Cummings\thanks{CONTACT Peter T. Cummings. Email: peter.cummings@vanderbilt.edu 
The content of this paper reflects that presented in the {\em Molecular Physics} lecture by Peter Cummings at the 2019 Thermodynamics conference in Punta Umbría, Huelva, Spain, June 26-28, 2019} }
\affil{Department of Chemical and Biomolecular Engineering, Vanderbilt University, Nashville, TN, USA \\ Multiscale Modeling and Simulation Center, Vanderbilt University, Nashville, TN, USA} }

\maketitle

\begin{abstract}
Systems composed of soft matter (e.g., liquids, polymers, foams, gels, colloids, and most biological materials) are ubiquitous in science and engineering, but molecular simulations of such systems pose particular computational challenges, requiring time and/or ensemble-averaged data to be collected over long simulation trajectories for property evaluation. 
Performing a molecular simulation of a soft matter system involves multiple steps, which have traditionally been performed by researchers in a ``bespoke'' fashion, resulting in many published soft matter simulations not being reproducible based on the information provided in the publications. To address the issue of reproducibility and to provide tools for computational screening, we have been developing the open-source \underline{Mo}lecular  \underline{S}imulation and  \underline{De}sign  \underline{F}ramework (MoSDeF) software suite.

In this paper, we propose a set of principles to create Transparent, Reproducible, Usable by others, and Extensible (TRUE) molecular simulations. MoSDeF facilitates the publication and dissemination of TRUE simulations by automating many of the critical steps in molecular simulation, thus enhancing their reproducibility. 
We provide several examples of TRUE molecular simulations: All of the steps involved in creating, running and extracting properties from the simulations are distributed on open-source platforms (within MoSDeF and on GitHub), thus meeting the definition of TRUE simulations.

\end{abstract}


\section{Introduction}\label{sec:intro}
Reproducibility in scientific research has become a prominent issue, to the extent that  some have opined that science has a ``reproducibility crisis''.\cite{Baker2016}
Along with the rest of the scientific community, computational scientists are grappling with the central question: How can a computational study be performed and published in such a way that it can be replicated by others?
This has become increasingly important as researchers seek to harness the ever expanding computational power to perform large-scale computational screening of materials\cite{Tadmor2013,Jain2013,Wilmer2012,Hachmann2011,Afzal2019,Thompson2019,Matsumoto2019} inspired by the materials genome initiative (MGI)\cite{OSTP2011}, where reproducibility issues commonly faced in small-scale studies will only be compounded as the number of simulations grow by orders of magnitude.

\begin{figure}
\centering
\includegraphics[height=5cm]{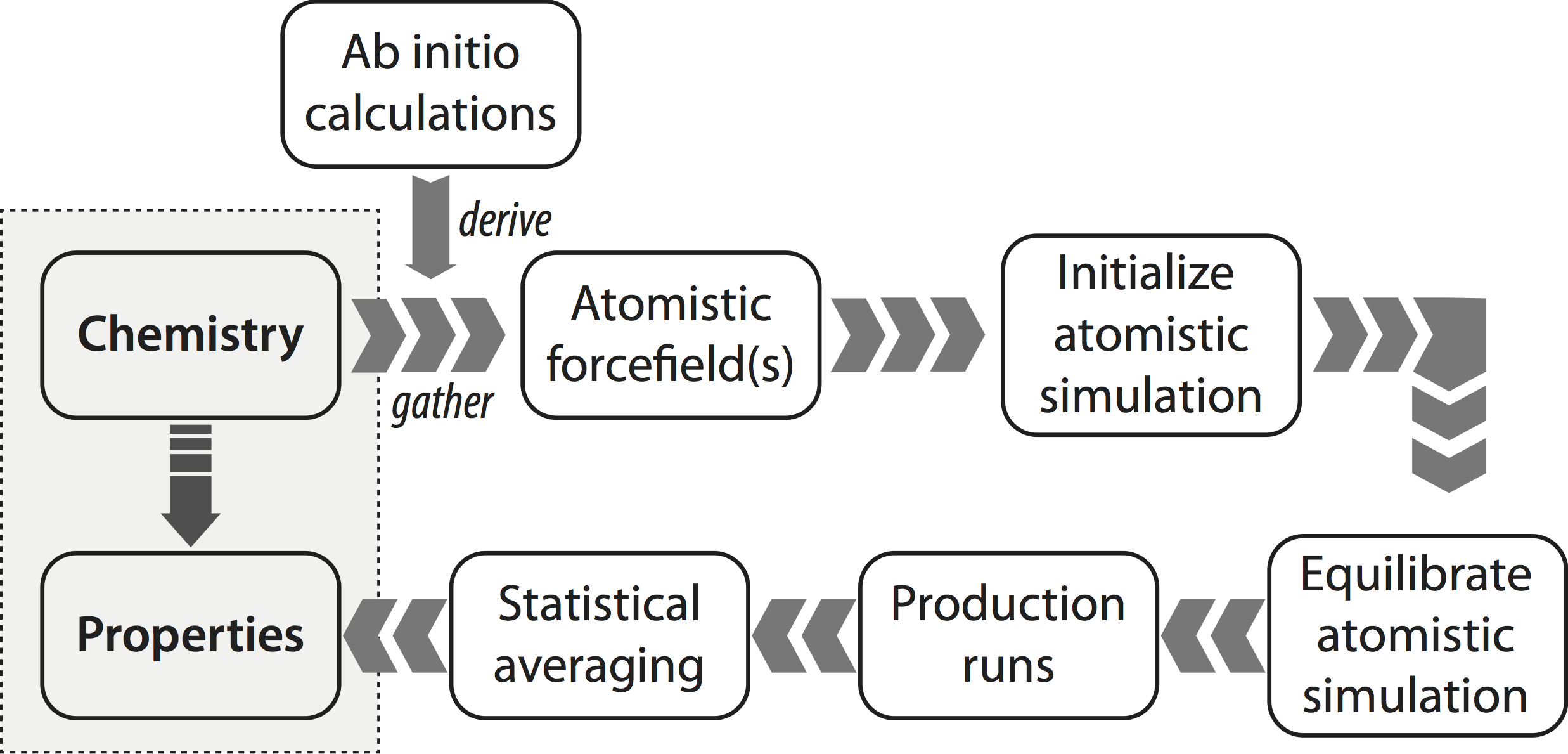}
\caption{
    Schematic of the typical process required to compute properties of soft matter systems from system ``chemistry,'' which refers to chemical composition and state (including temperature, pressure and composition), starting from the need to either gather or derive force field parameters to model the system.
    For coarse-grained (CG) simulations, the CG force fields are often derived from atomistic simulations.
}
\label{fig:molsimflow}
\end{figure}

Addressing the issues of reproducibility in soft matter simulation is particularly challenging, given the complexity of the simulation inputs and workflows.
Here we define soft matter as anything easily deformed at room temperature, e.g., liquids, polymers, foams, gels, colloids, and most biological materials.
Fig.~\ref{fig:molsimflow} shows a schematic of the general multi-step workflow for performing atomistic simulations of soft matter systems, proceeding from system ``chemistry'' (chemical composition and state conditions such as phases(s), temperature, pressure, and composition) to ``properties'' (e.g., structural, thermodynamic and transport properties, phase equilibria, and dielectric properties) In such systems, the differences in potential energy between distant configurations are on the same order as the thermal motion, requiring time and/or ensemble-averaged data to be collected over long simulation trajectories for property evaluation.
The equilibration procedures and system sizes considered may strongly influence the resulting measured properties, since one must consider both the local conformations of the underlying components, along with any mesoscopic structuring present in the system.
To capture sufficiently large length and time scales, soft matter simulations are typically performed using methods such as molecular dynamics (MD) or Monte Carlo (MC) that employ empirical force fields to model the interactions between atoms and molecules; the appropriate force field parameters must be identified before the simulation can be performed.

Some commonly available force fields, such as the Optimized Potential for Liquid System (OPLS)\cite{Jorgensen1996d} and the General Amber Force Field (GAFF)\cite{Wang2004} contain thousands of possible parameters that are differentiated by their chemical context (e.g., the element a given interaction site represents, the number and identity of bonded neighbors, the local environment of bonded neighbors, the type of system, etc.).
Selecting the appropriate force field parameters for a particular use case is often non-trivial.
Workflows may also involve the optimization of specific parameters, such as partial charges, or require separate procedures to develop novel force fields, such as coarse-grained (CG) models, before a simulation can be performed.
Furthermore, due to the complex nature of the underlying constituents (e.g., highly branched polymers), setup of an initial system configuration may be challenging and require additional relaxation procedures to ensure system stability\cite{Jo2007}.

As such, soft matter simulations typically require multi-step workflows with many inputs.
The various steps are often accomplished by separate pieces of syntactically and/or semantically incompatible software tools, that may require translators or {\em ad hoc} modifications to facilitate interoperability.
These tools, and especially the ``glue'' code that facilitates interoperability, are typically neither publicly available nor version-controlled.
The latter is particularly important for long-term reproducibility, since to repeat a particular calculation may require using versions of the relevant codes used when the work was originally published, which could be a number of years ago.

The above complexities often make it difficult for researchers themselves to fully capture and preserve the procedures used to perform a simulation, let alone clearly disseminate these to the rest of the community.
A typical soft matter simulation publication provides an overview of the methods and procedures used but falls significantly short of including the necessary information to unambiguously reproduce the published work.
This information includes, but is not limited to, citations to the sources of force field(s) used, the numeric parameters of the force field(s) used, how the force field parameters were assigned to the system, constants and options provided in the underlying algorithms, and the exact choices used in constructing the initial configuration of the system.
It is important to recognize that the results from a simulation can depend on the minute details\cite{Schappals2017}.
These details include, but are not limited to, the random seed used to generate a distribution, the specific force field parameters and how they were used, the exact procedures employed to equilibrate a system, etc.
For example, small variations in force fields (e.g., changes in distances at which interactions are truncated, different partial charges, the specific method for handling long-ranged interactions, etc.) can change some predicted properties quite significantly\cite{Chen2003,SilvaFernandes2004,Schappals2017,SilvaFernandes2004}.  The minute details may also be inherent to the software used to perform the simulations, and thus the use of ``in-house'' or commercial (i.e., closed-source) software stymies reproducibility.
If the source code cannot be viewed, the underlying algorithms and inputs cannot be examined, the quality of the code and whether it has undergone proper validation is unknown, and errors cannot be identified.
As an example, a long standing disagreement related to phase transitions in supercooled water was only recently settled after the source code of the in-house software used to perform the calculations was shared.
The differences in observed transitions were attributed to a subtle error in how velocities were assigned when initializing the many short MD simulations in the hybrid MD/MC workflows.\cite{Palmer2018,Smart2018}.
The use of open-source simulation engines therefore clearly enhances reproducibility, as the underlying source code can be examined (note, the use of open-source simulation engines is now routine for MD studies, but often these engines and other open-source codes are modified to implement new force field parameters or functional forms, and MC studies still commonly use in-house software).
However, it is atypical for input scripts and data files for open-source simulation engines to be included as part of a publication and thus reproducibility still largely depends on the thoroughness of the description of the methods and model in the text.
Furthermore, the algorithms and specific choices used to generate a data file, which may influence the results and their validity (e.g., how a force field was applied), are lost if the software and/or procedures used to generate the data file are not made available.
Even when using open-source simulation engines, researchers still routinely use in-house software for other steps in the process, i.e. generation of initial configurations, selection of force field parameters, and analysis.
Furthermore, if a workflow relies upon manipulation or modification of individual pieces of software by a user (e.g., initializing a system using software with a graphical user interface, GUI\cite{Jo2008}), or human-modification of files, it is often difficult to capture and convey the exact procedures in such a manner that they can be reproduced by another researcher.

Fortuitously, several researchers have proposed general guidelines for increasing reproducibility in computational research, which can be used to infer best practices for soft matter simulation.
Donoho {\em et al.}\cite{Donoho2009} propose that all details of computations -- code and data -- should be made ``conveniently available'' to other researchers; they also provide arguments in favor of the creation and use of community developed software libraries and the use of scripting.
Others\cite{Sandve2013,Elofsson2019} have proposed succinct ``rules'' as keys to reproducible computational research, including version control, replacement of manual input with scripts, and public access to these scripts, input files, and resulting data.
It was also noted that computational frameworks that integrate different tools within a common environment naturally satisfy many of these rules.
One of the most vocal proponents of reproducibility in computational science\cite{Barba2016,Barba}, has gone as far as asserting that GUIs are the ``enemy of reproducibility''.
GUIs hide details and require human interaction and manipulation in contrast to scripts, which fully reveal the way in which calculations are performed.
A classic example is Excel spreadsheets, where the relationship between calculation cells and data is normally hidden, and the order of calculation is not obvious, nor necessarily controllable.
In 2010, Harvard University economists Reinhart and Rogoff published a highly cited and influential paper on the role of debt in limiting growth in national economies\cite{Reinhart2010}.
The study, based on data manipulated within an Excel spreadsheet, was often cited by politicians favoring austerity policies in the wake of the 2008 financial crisis while public economic policy was being formed.
Subsequently, Herndon {\em et al.}\cite{Herndon2014} found that the spreadsheet contained errors in formulae that dramatically changed the conclusions.

Determining how these guidelines for reproducibility should be –- and/or can be –- implemented in soft matter simulation is in itself a challenge.
For example, simply providing code is not effective if that code is poorly written or not well documented and has subtle issues, such as dependencies within a code (e.g., use of external libraries, especially if they are proprietary/non-free or difficult to obtain/install)
These issues may create barriers to proper compilation/installation and hence hamper reproducibility.
Similarly, providing a raw data file without defining the structure of it, and/or without appropriate metadata, does little to aid in reproducibility.
Since journals largely do not provide mechanisms for sharing code, scripts, and/or data (aside from supplemental material), it is also not clear how such information should best be shared.

As such, in order to implement best practices, we assert that the development of new tools and standards will be required, in order to facilitate necessary changes to the way in which simulators perform and publish their research.
However, development of new tools does little to improve reproducibility if those tools are not used; to be widely adopted by the community, they must provide additional value to researchers, e.g., minimizing errors, reducing development time, preventing knowledge loss, providing novel functionality, etc.

For almost a decade now, we have been developing a robust Python-based, open-source integrated software framework for performing simulations of soft matter systems with the goal of implementing best practices and enabling reproducibility.
This framework, known as the \underline{Mo}lecular  \underline{S}imulation and  \underline{De}sign  \underline{F}ramework (MoSDeF)\cite{mosdefwebpage}, was developed initially at Vanderbilt University, in collaboration with computer scientists in the Institute for Software Integrated Systems\cite{ISIS}, to facilitate screening studies of monolayer lubrication using MD methods.
MoSDeF provides a core foundation and includes tools for programmatic system construction (\texttt{mBuild})\cite{Klein2016b,mbuildgithub}, tools for encoding force field usage rules and their application (\texttt{Foyer})\cite{foyergithub,Iacovella2016c,Black2017a}, and has recently integrated the \texttt{signac} framework\cite{Adorf2018,signacdocs}, developed at the University of Michigan as a means of improved data and workflow management.
The MoSDeF toolkit has been used in various published results\cite{Klein2016b,Black2017a,Summers2017a,Thompson2019,Matsumoto2019,Summers20} and ongoing research projects, with the primary MoSDeF tools having each been downloaded over 18,000 times from Anaconda Cloud\cite{mosdefanaconda} since February 2017.
Despite being initially developed for monolayer lubrication, the underlying tools can be and have been applied to soft matter systems in general, and the modular, object-oriented design naturally allows for intuitive extension.
Current MoSDeF activities are expanding the capabilities related to:
\begin{itemize}
\item Initializing system configurations by providing a plugin architecture for community contributions
\item Providing initialization routines for a wide variety of common systems
\item Developing an improved backend that will support an increased number of force field types and simulation engines, including open-source MC software
\item Developing modules that implement methods for partial charge assignment
\item Including improved support and libraries for coarse-grained models
\item Developing modules that allow for reproducible derivation of coarse-grained and atomistic force fields
\item Developing workflows for free energy methods and phase equilibria
\item Specifically identifying and implementing best practices within the various modules/workflows that improve reproducibility.
\end{itemize}
Through the MoSDeF integrated framework, the exact procedures used to set up and perform simulation workflows and associated metadata (i.e., the provenance) can be scripted, encapsulated, version-controlled, preserved, and later reproduced by other researchers.
This allows molecular simulation studies to be conducted and published in a manner that is TRUE: Transparent, Reproducible, Usable by others, and Extensible.

The remainder of this paper is organized as follows.
In Section~\ref{sec:mosdef}, we briefly review MoSDeF an its capabilities. In Section~\ref{sec:TRUEmolsim}, we consider four examples of TRUE molecular simulations in diverse application areas.
Finally, in Section~\ref{sec:conclusions}, we summarize our conclusions and prospects for future development of MoSDeF.

\section{Overview of MoSDeF}\label{sec:mosdef}
\subsection{MoSDeF tools and capabilities}\label{tools}
MoSDeF is a set of an open-source Python libraries, designed to facilitate the construction and parameterization of systems for molecular simulation. MoSDeF includes routines to output syntactically correct configuration files  in formats used by common simulation engines, currently supporting  
GROMACS\cite{Hess2008,Berendsen1995,Abraham2015}, LAMMPS\cite{Plimpton1995a}, HOOMD-blue\cite{Anderson2008a,HOOMDWeb}, and Cassandra \cite{Shah2017a},   as well as  other common file formats (e.g., \texttt{MOL2}, \texttt{PDB}) through integration with the open-source ParmEd\cite{ParmEd} parameter editing package.
Each library (i.e., Python module) in MoSDeF is designed such that it can be used as a standalone package, in combination with other libraries within MoSDeF, or with other libraries developed and used by the community.
This composability/modularity is an essential design feature in terms of the robust development of MoSDeF,  allowing the framework to be more modifiable, testable, extensible, and have fewer bugs than monolithic approaches\cite{Schappals2017}.
MoSDeF is implemented using concepts from the computer science/software engineering field of model integrated computing (MIC)\cite{Sztipanovits:1997ve,Iacovella2012a}, a systems engineering approach, pioneered at the Institute for Software Integrated Systems (ISIS) at Vanderbilt, that emphasizes  the creation of domain-specific modeling  languages that capture the  features of the individual components of a given process, at the appropriate level of abstraction.
By using concepts from MIC, MoSDeF can easily be abstracted and is able to capture the relationships that exist between data and processes regardless of the level of abstraction, essential for ensuring that system initialization scripts are transparent and usable by others.
MoSDeF follows a modern open-source development model with special emphasis on effective code sharing, accepting external feedback, and bug reporting.
\begin{itemize}
\item All modules and workflows developed for MoSDeF build upon the scientific Python stack, thus enabling transparency, promoting code reuse, lowering barriers to entry for new users, and promoting further community driven, open-source development.
\item GitHub is used for hosting MoSDeF's version-controlled software development, deployment, and documentation/tutorials, using a pull request (i.e., fork-pull) model that allows for code review and automated testing, helping ensure proper standards have been followed and allows for automated testing of software and software artifacts.
\item Automated builds and testing of the software are hosted on Travis CI\cite{TravisCI} and also on Microsoft's Azure Pipelines\cite{AZP} to ensure that proposed modifications to the code do not break the current performance and the \texttt{CodeCov}\cite{CodeCov} tool is used to ensure that modifications to the code are covered by unit tests.
\item All software developed as part of the MoSDeF project are open-source, with the standard MIT license\cite{MIT} that allows free use, reuse, modifications, as well as commercialization.
\item Slack\cite{Slack} is used to facilitate effective collaborative communication and software development across a wide geographic area\cite{Davenport2016}.
\end{itemize}

By developing software in a modular, extensible, open-source manner, using freely available tools designed for collaborative code development (e.g., git, GitHub, and Slack), we are creating a long-term community-developed effort, similar to the success seen by other tools in the community (e.g., GROMACS\cite{Hess2008,Berendsen1995,Abraham2015}, VMD\cite{Humphrey1996}, LAMMPS\cite{Plimpton1995a}, HOOMD-Blue\cite{Anderson2008a,HOOMDWeb}, etc.).
This has become especially important as the group of MoSDeF developers has expanded  beyond Vanderbilt University.
A recent U.S. National Science Foundation grant\cite{NSFMoSDeF} has provided support for leading molecular simulation research groups from Vanderbilt, the universities of Michigan, Notre Dame, Delaware, Houston and Minnesota, along with Boise State University and Wayne State University to further improve and increase support of MoSDeF as described below.
This group spans a broad range of expertise, and an equally broad range of scientific applications, open-source simulation codes (HOOMD-Blue\cite{Anderson2008a,HOOMDWeb}, Cassandra\cite{Shah2017a}, GOMC\cite{GOMChome,GOMC2018} and CP2K\cite{CP2Khome}), workflow and data management software\cite{Adorf2018,signacflowdocs,signacdocs}  and algorithms and analysis tools; computer scientists from ISIS are also involved in the collaboration, helping to ensure the use of best practices and provide novel insight into algorithmic and software development.
In combination, this collaboration is working to dramatically expand the capabilities of MoSDeF and thus facilitate researchers in the area of molecular simulation to be able to publish TRUE simulations.

Here, we briefly describe the two key tools used in the current version of MoSDeF, focusing on the specific aspects of the tools that help to enable TRUE simulations.

\subsubsection{mBuild}\label{sec:mbuild}
The \texttt{mBuild} Python library\cite{Klein2016b,mbuildgithub} is a general purpose tool for constructing system configurations in a programmatic (i.e., scriptable) fashion.
While tools exist in the community for system construction\cite{Salomon-Ferrer2013a,Plimpton2006,Jewett2013}, they tend to be system specific (e.g., bilayer construction),  often employ GUIs which may hamper reproducibility\cite{Barba2016} and may be limiting for workflows that require automation,  and most are designed around the concept that components of the system can be described by self-contained templates  (e.g., where a system can be constructed by simply duplicating a template that describes a molecule).
Such existing tools have typically not been designed to work for systems where bonds are added between different components (e.g., polymer grafted surfaces)  or for systems where one component is semi-infinite (e.g., a silica substrate that is periodic in-plane)  and most do not allow programmatic variation of specific structural/chemical aspects (e.g., the length of a polymer, the polymer repeat unit, size of a substrate, etc.);  \texttt{mBuild} was designed specifically to provide this missing functionality.  

Rather than providing a tool to perform initialization that only applies to a specific family of systems (e.g., monolayers), \texttt{mBuild} provides  a library of functions that users can combine, extend, and add to, in order to construct specific systems of interest.
\texttt{mBuild} allows users to hierarchically construct complex systems from smaller,  interchangeable pieces that can be connected, through the use of the concept of generative, or procedural, modeling\cite{Klein2016b}.
This is achieved through \texttt{mBuild}'s underlying \texttt{Compound} data structure, which is a general purpose ``container'' that can describe effectively anything within the system:  an atom, a coarse-grained bead, a collection of atoms, a molecule, a collection of \texttt{Compound}s,  or operations (e.g., a  \texttt{Compound} that includes a routine to perform polymerization).
To  join \texttt{Compound}s (e.g., attachment of a \texttt{Compound} that defines a polymer  to a \texttt{Compound} that defines a surface),  \texttt{Compound}s can include \texttt{Port}s that define both the location and orientation of a possible connection.
In \texttt{mBuild}, a user (or algorithm) defines which \texttt{Port}s on two \texttt{Compound}s should be connected  and the underlying routines in the software automatically performs the appropriate translations and orientations of the \texttt{Compound}s  (see Klein {\em et al.}\cite{Klein2016b} for more details).
This creates a new (composite) \texttt{Compound} that contains both of the original \texttt{Compound}s, now appropriately oriented and positioned in space,  with an explicit bond between them;  since \texttt{Compound}s are general data structures, the same operations  (rotation, translation, connecting of \texttt{Port}s, etc.) that were performed on the underlying \texttt{Compound}s  can be performed on this  new composite \texttt{Compound}.
The \texttt{mBuild} library can be used to create systems from ``scratch'' whereby a user implements all the  relevant code to define the building blocks and how they should be connected, or by using and/or extending the various``recipes''  included in \texttt{mBuild}.  
\texttt{mBuild} includes (but is not limited to)  ``recipes''  for initializing polymers, tilings (e.g., duplicating a unit cell, including bonding information), patterning (disk, sphere, random, etc.), lattices either from a Crystallographic Information File (CIF), their Bravais lattice parameters, or the vectors describing the prism, box filling (via integration with PACKMOL\cite{packmol}), monolayers and brushes on flat, curved, and spherical surfaces, and bilayers and lamellar structures.  

\begin{figure}
\centering
\includegraphics[height=10cm]{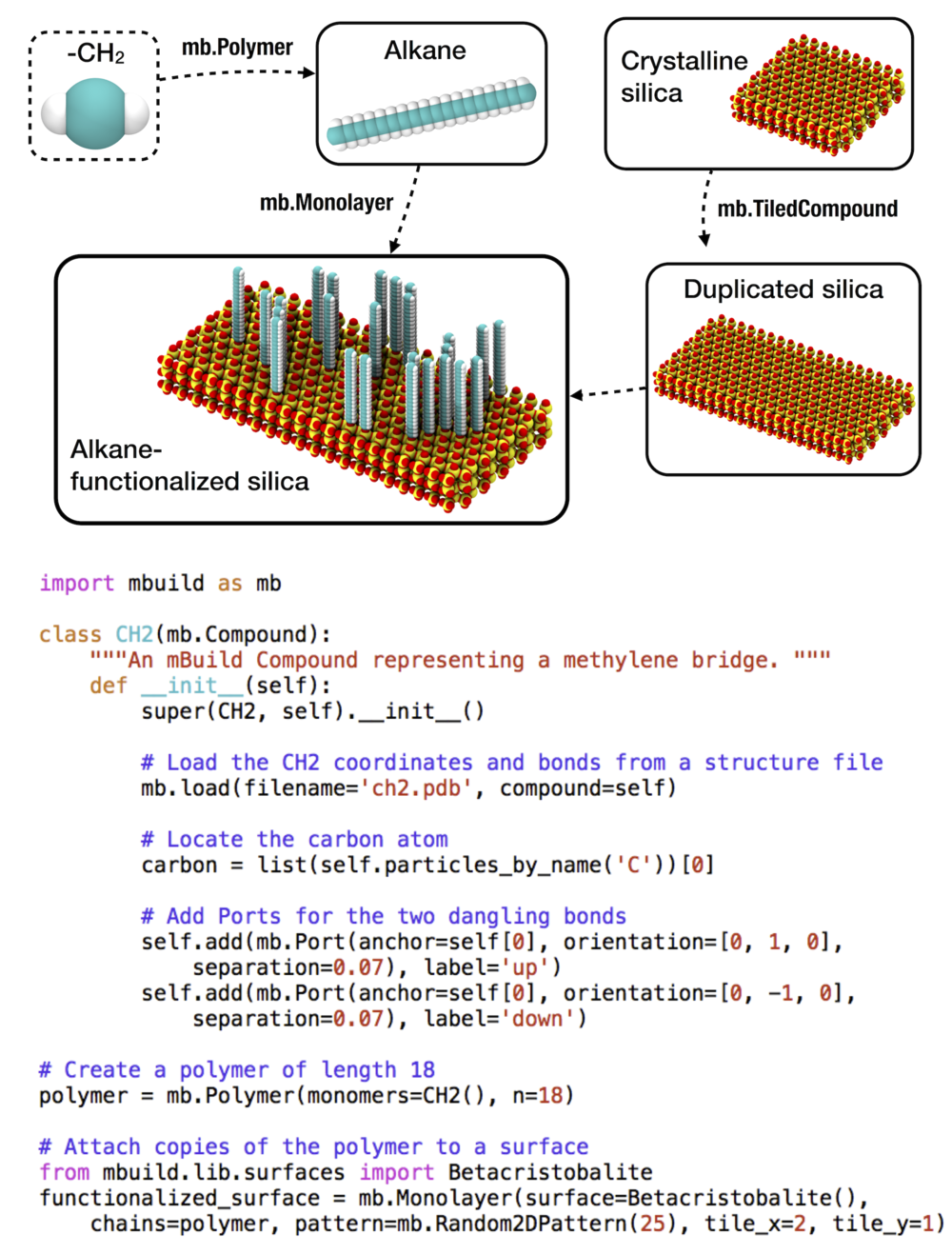}
\caption{
    Python script that uses \texttt{mBuild} to define a class for a --CH$_2$-- group, create a polymer composed of multiple --CH$_2$-- groups, and connects copies of this polymer to a surface.
    Note for simplicity, the terminal CH$_3$ group is not shown.
}
\label{fig:mbuild-listing}
\end{figure}

As an example, Fig.~\ref{fig:mbuild-listing} shows a schematic and associated code for the construction of an alkane grafted silica surface.
In this code, a custom \texttt{Compound} class is defined for a CH$_2$ moiety alongside a \texttt{Compound} from the  \texttt{mBuild} library that defines a crystalline silica surface; a``recipe'' included in \texttt{mBuild} that performs polymerization (\texttt{Polymer}) is used to connect copies of the CH$_2$ moieties; the \texttt{Monolayer} ``recipe'',  also included in \texttt{mBuild}, is used to perform the functionalization of the silica surface with the polymer,  returning a single composite  \texttt{Compound}  of the functionalized surface (for readability, the terminal groups are ignored in this example).
This example also highlights how system construction can be programmatically varied, e.g., the \texttt{Polymer} class takes as input the number of repeat units (in this case, set to 18). Similarly, the size of the substrate can be toggled in the  \texttt{Monolayer} class, where \texttt{tile\_x} and \texttt{tile\_y} define the number of times the substrate is duplicated in the respective dimension.
The number of chains attached to the surface  can also be modified via the number passed to the   \texttt{Random2DPattern} class (here set to 25).
Because \texttt{Compound}s are general containers, changes to, e.g.,  the length of the polymer, do not require changes to the rest of the script, namely the \texttt{Monolayer} class.  
Similarly, characteristics such as the repeat unit passed to the \texttt{Polymer} class can be readily changed without need to change other aspects of the script.   
As a result, by using the \texttt{mBuild} library, complex system initialization  and variation/extension can often be accomplished without the need to write significant amounts of code.
As this example shows, by using concepts from MIC, construction of systems in \texttt{mBuild} can be trivially abstracted  (i.e.,  the level of complexity reduced) to the level most appropriate to describe (i.e., model) the system, without making system construction a ``black box''.
Since \texttt{mBuild} is an open-source, freely available Python library, scripts that unambiguously define all the steps needed to initialize a system can be easily shared and disseminated with publications, with all code easily interrogated,  allowing system construction to be reproduced and improving transparency; 
 \texttt{mBuild} has additionally been architected so that users can contribute custom ``recipes'' for system initialization via a plug-in environment, further allowing such routines to be easily shared,  utilized and extended by others.

\subsubsection{Foyer}\label{sec:foyer}
The \texttt{Foyer} library\cite{foyergithub} is a tool for applying force fields to molecular systems (i.e., atom-typing).
\texttt{Foyer} provides a standardized approach to defining chemical context (i.e., atom-typing rules)\cite{Iacovella2016c,Klein2019} along with the associated force field parameters.
While there are freely available tools to aid in atom-typing\cite{Wang2006b,Vanommeslaeghe2012a,Malde2011a,Ribeiro2008a,Eggimann2014},  these are typically specific to a particular force field or simulator, and/or capture the atom-typing and parametrization in a hierarchy (either through specific placement in a parameter file read by the code or as nested if/else statements within the source code).  
\texttt{Foyer} does not encode usage rules into the source code, instead defining usage rules and parameters in an \texttt{XML} file that is an extension of the \texttt{OpenMM}\cite{OpenMM} force field file format.
The \texttt{Foyer} software itself is used to interpret and apply the rules and thus the software is not limited to use with only a single force field type.
By separating the usage rules from the source code, changes or extensions to force field parameters/rules does not require changes to the code itself.
Force field usage rules are encoded using a combination of a SMARTS-based\cite{SMARTSweb} annotation scheme, which defines the molecular environment  (i.e., chemical context) associated with a given parameter, and \texttt{overrides} that define rule precedence (i.e., which atom type to choose when multiple rules can apply to an interaction site).
The use of \texttt{overrides}  avoids the need to define rule precedence via the order of the rules within a file (See \cite{Klein2019} for more details).
As an example, Listing 1 shows the contents of an \texttt{XML} file that contains parameters and usage rules from the OPLS force field  for linear alkanes.  
We note that \texttt{Foyer} allows user-defined input (by pre-pending with an underscore), allowing SMARTS to be used  for non-elemental interaction sites (e.g., an interaction site that represents a coarse-grained bead or an united atom interaction site).
As such, the exact parameters and their usage can be readily captured and disseminated along with a simulation and/or publication.
This provides an improved way to disseminate custom force field parameter sets and/or novel force field parameters (e.g., see Ref. \cite{Black2017a}) that reduces ambiguity, as the format used by \texttt{Foyer} to encode the usage rules and parameters is both human and machine readable; thus parameterization rules provided in a publication can be automatically tested for accuracy and completeness.
To further  enhance reproducibility, the \texttt{XML} force field files additionally include a \texttt{doi} tag for the source of the parameters (see  Listing 1); upon successful atom-typing, \texttt{Foyer} outputs a BibTeX file of references with the relevant DOIs, significantly improving the transparency as to the origin of parameters used in a simulation and therefore reproducibility.   

\begin{lstlisting}[basicstyle=\tiny\ttfamily]
<ForceField>
  <AtomTypes>
    <Type name="opls_135" class="CT" element="C" mass="12.01100" def="[C;X4](C)(H)(H)H" 
    desc="alkane CH3" doi="10.1021/ja9621760"/>
    <Type name="opls_136" class="CT" element="C" mass="12.01100" def="[C;X4](C)(C)(H)H" 
    desc="alkane CH2" doi="10.1021/ja9621760"/>
    <Type name="opls_140" class="HC" element="H" mass="1.00800" def="H[C;X4]" 
    desc="alkane H" doi="10.1021/ja9621760"/>
  </AtomTypes>
  <HarmonicBondForce>
    <Bond class1="CT" class2="CT" length="0.1529" k="224262.4"/>
    <Bond class1="CT" class2="HC" length="0.1090" k="284512.0"/>
  </HarmonicBondForce>
  <HarmonicAngleForce>
    <Angle class1="CT" class2="CT" class3="CT" angle="1.966986067" k="488.273"/>
    <Angle class1="CT" class2="CT" class3="HC" angle="1.932079482" k="313.800"/>
    <Angle class1="HC" class2="CT" class3="HC" angle="1.881464934" k="276.144"/>
  </HarmonicAngleForce>
  <RBTorsionForce>
    <Proper class1="CT" class2="CT" class3="CT" class4="CT" c0="2.9288" c1="-1.4644" 
    c2="0.2092" c3="-1.6736" c4="0.0" c5="0.0"/>
    <Proper class1="CT" class2="CT" class3="CT" class4="HC" c0="0.6276" c1="1.8828" 
    c2="0.0" c3="-2.5104" c4="0.0" c5="0.0"/>
    <Proper class1="HC" class2="CT" class3="CT" class4="HC" c0="0.6276" c1="1.8828" 
    c2="0.0" c3="-2.5104" c4="0.0" c5="0.0"/>
  </RBTorsionForce>
  <NonbondedForce coulomb14scale="0.5" lj14scale="0.5">
    <Atom type="opls_135" charge="-0.18" sigma="0.35" epsilon="0.276144"/>
    <Atom type="opls_136" charge="-0.12" sigma="0.35" epsilon="0.276144"/>
    <Atom type="opls_140" charge="0.06" sigma="0.25" epsilon="0.12552"/>
  </NonbondedForce>
</ForceField>
\end{lstlisting}
Listing 1. \texttt{OpenMM} formatted \texttt{XML} file for linear alkanes using the OPLS force field\cite{Jorgensen1996d}.

\subsection{Other Community Tools}\label{sec:othertools}

Here we briefly highlight other simulation tools and efforts with a considerable focus on reproducibility and transparency, several with similar and/or complementary functionality to MoSDeF.  
We do not include discussion of commercial tools, as the need to purchase software places a fundamental roadblock in terms of reproducibility.

The Atomic Simulation Environment (ASE)\cite{HjorthLarsen2017} is a Python toolkit that provides wrappers to various programs/libraries allowing atomistic simulations to be setup, run and analyzed within a single script.
Support is provided for numerous electronic structure codes and several MD simulation engines; however, as ASE is primarily focused on hard matter systems it does not currently support robust tools for initialization of complex soft matter systems or atom-typing.

\texttt{Pysimm}\cite{Fortunato2017,Fortunato}, is an open-source Python toolkit for soft matter systems providing routines for system setup and wrappers that support LAMMPS MD\cite{Plimpton2006} and Cassandra MC\cite{Shah2017a} engines, allowing a simulation workflow to be encoded in a Python script.
Of particular note, \texttt{pysimm} includes routines that simplify the process for performing complex workflows such as simulated growth/crosslinking of polymers\cite{Abbott2013}.
We note that since both ASE and \texttt{pysimm} are also developed as Python libraries, there is a natural level of interoperability between these tools and  MoSDeF.   
\texttt{Hoobas} is another open-source molecular building package that facilitates the construction of polymers for molecular dynamics simulation \cite{Girard2019,hoobasgithub}.
\texttt{indigox} is an open-source package that utilizes the CherryPicker algorithm to help parametrize molecules based on fragments of previously-parametrized molecules \cite{indigoxgithub}.
\texttt{Open Babel} is a library of cheminformatics functions that support constructing molecular models, SMARTS-matching, and basic molecular dynamics functions with basic molecular mechanics force fields \cite{OBoyle2011,openbabelgithub}.
\texttt{OpenKIM} is a multifaceted toolkit providing a portal for storage of interatomic models and  their associated data, and an application programming interface (API) created such that models can work seamlessly (and correctly) between different simulation engines;  we note this API is designed to ensure parameters are defined correctly, not to perform atom-typing or to encode usage rules and does not provide tools for system initialization or workflow management.
To date, \texttt{OpenKIM} has largely focused on atomic systems (i.e., a system is defined solely by its atoms, and “bonds” are an outcome of atomic positions), whereas most soft-matter force fields include both non-bonded and bonded parameters and assign different parameters to atoms based on the bonds.
The Open Force Field consortium\cite{Mobley2018,Zanette2018,david_l_mobley_2016_154235} has developed a variety of open-source tools  that utilize chemical perception via SMIRKS\cite{SMIRKS} patterns to identify atom types and other force field parameters pertinent to each atom in a chemical system, similar to \texttt{Foyer}'s underlying methodology.
\texttt{WebFF} is an ongoing NIST-project aimed at developing an infrastructure for modeling soft materials and curating force field data for traceable data provenance \cite{webffgithub}.
\texttt{BioSimSpace} provides an API that allows users to mix-and-match various molecular modeling tools, facilitating the use of complex workflows involving molecular dynamics, metadynamics, various water models, various force fields, free energy methods, and various simulation engines\cite{Hedges2019}.
\texttt{signac} is a Python library that provides basic components required to create a well-defined and collectively accessible data space and enables data access and modification through a homogeneous data interface that is agnostic to the data source. \texttt{signac-flow} is an extension of the \texttt{signac} framework\cite{signacdocs}, which aids in the management of highly complex data spaces. \texttt{signac-flow} allows submission to high performance computing (HPC) scheduling systems, including both PBS and SLURM.
Since \texttt{signac-flow} captures the entire workflow definition and execution, it can be used to facilitate reproducible workflows.
 \texttt{mBuild} and \texttt{Foyer} have been used in combination with \texttt{signac-flow} in several past and on-going research projects by the authors\cite{Thompson2019,Matsumoto2019}.  
\texttt{FireWorks} is another workflow manager that supports dynamic workflows using \texttt{MongoDB}\cite{Jain2015,fireworksgithub}.

\section{TRUE Molecular Simulations}\label{sec:TRUEmolsim}
We have defined TRUE molecular simulations as ones that are \underline{t}ransparent, \underline{r}eproducible, \underline{u}sable by others and \underline{e}xtensible.
In this section, we provide some examples of TRUE simulations utilizing the capabilities of MoSDeF. But first we define what we mean by these terms in the context of molecular simulation.

A simulation is  {\em transparent} when all the information needed to exactly
follow the steps undertaken by the original author(s) (such as all scripts used to set up the system, details of force field implementation, all input files to the simulation engines, any other needed input files) are visible to anyone in the community. This requires the sharing of this information in a version-controlled persistent open-source repository, such as GitHub.
This information, in and of itself, may only be useful to a true expert; however, few simulations published today meet even this standard.
A transparent simulation is {\em reproducible} when sufficient information (in supplementary information and/or documentation) is provided so that future researchers interested in duplicating the work can construct and run the reported simulation.
From this point of view, a self-contained workflow - such as a Jupyter notebook, or a virtual machine - is highly desirable. As defined here, reproducibility does not require a high level of expertise - for example, the calculation could be reproduced by a student in a class, a newcomer to simulation, etc.
In particular, Jupyter notebooks provide a convenient, high-level representation of a script that integrates with other common Python tools and can be converted directly into a Python script using \texttt{nbconvert}.
Two caveats about reproducibility must be borne in mind.
First, we note that in molecular simulation, reproducibility is unlikely to be exact, in the following sense:
Two MD simulations, when run on different architecture machines, will not generate the same trajectory due to differences in the handling of floating point operations.
As in any nonlinear dynamical system, small differences between trajectories (due to different rounding errors) grow exponentially large over time.
Even when run on the same computer, two simulations may not give the exact same trajectory.
This is because of parallelized computing, in which parts of the calculation are done by separate processors and then gathered (added together) in an order that is not predictable due to fluctuations in message passing times.
The problem is exacerbated even more in MC simulations, where a difference in random number seed will generate a different sequence of random numbers on the same machine with the same random number generator.
On different machines, trying to achieve reproducibility in MC simulations at the level of configurations on different machines requires using the same random number generator with reproducible arithmetic (IEEE standard-compliant) with the same seed; additionally, the same issue with parallelization noted for MD simulations also applies.\cite{Phillips2011}
However, we do not expect reproducibility at the level of individual simulation trajectories; what we expect is statistical reproducibility in the averages of the properties calculated over the course of the simulation.
Second, many simulations that are reported in the literature require prodigious amounts of computational resources, such as millions of hours on a leadership-class supercomputer.
In this case, having available all of the codes means that reproducibility is limited to those having available to them similar levels of computing resources.
In this case, we propose that researchers may also elect to make available a simplified version of the reported calculation accessible to those that have limited computational resources (for example, using a much smaller system size and shorter simulation times or a single physiochemical statepoint instead of many).
Such scaled-down versions could also have considerable pedagogical value.

We define a transferable, reproducible simulation to be {\em usable by others} when a future researcher can utilize the available files and documentation to reproduce the calculation and make use of the data generated - for example, to analyze the trajectory/trajectories for different properties.
This requires a level of documentation that includes information about where output files are located in the data space created by reproducing the simulation, and how these files might be analyzed in different ways.
Finally, a transferable, reproducible, usable-by-others simulation is {\em extensible} if the documentation is sufficiently detailed that a future researcher could change characteristics of the simulation - such as changing molecular species, state conditions, simulation engine, properties calculated, etc.

By adhering to the principles of TRUE simulations, researchers will enable their work to be utilized in ways that have not been hitherto possible.
In particular, it will create resources that lower the barrier to entry into the field of molecular simulation, as well as allow researchers to distribute their research in a more useful fashion.
Using MoSDeF is not necessary to create TRUE simulations, but as the examples below illustrate, MoSDeF makes it considerably easier to distribute TRUE simulations by automating and standardizing many of the steps, thus minimizing the documentation needed to create a TRUE simulation.
Also, the open-source nature of MoSDeF offers the ability for researchers to make contributions to the code base in the form of methods, recipes, force fields, etc.

\subsection{Ethane VLE using TraPPE}
\begin{figure}[H]
    \includegraphics[width=\textwidth]{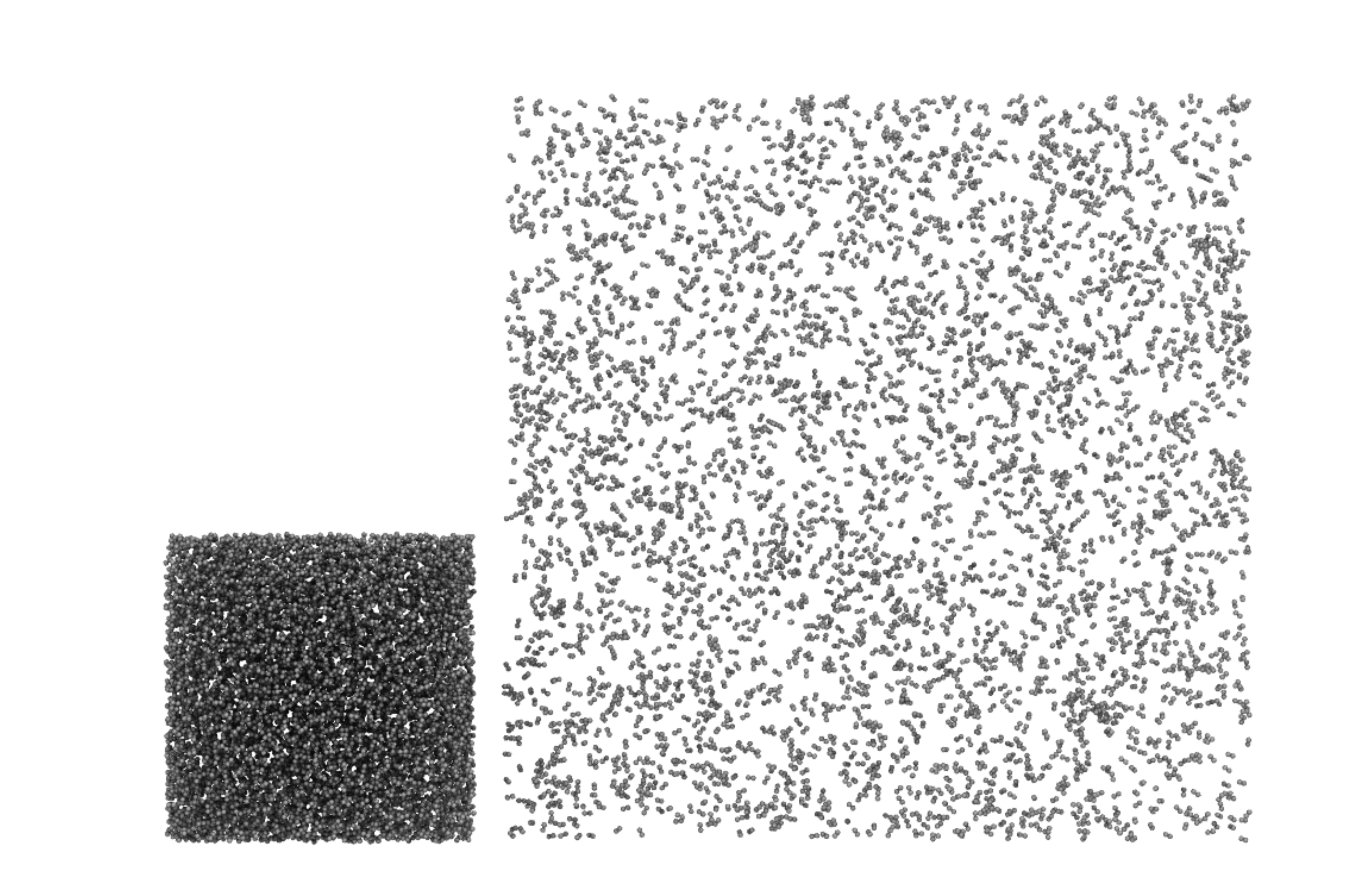}
    \caption{Two boxes of ethane constructed in \texttt{mBuild}.  Liquid phase (left) and vapor phase (right) are simulated simultaneously in GEMC.}
    \label{fig:vleboxes}
\end{figure}

\begin{figure}[H]
    \includegraphics[width=\textwidth]{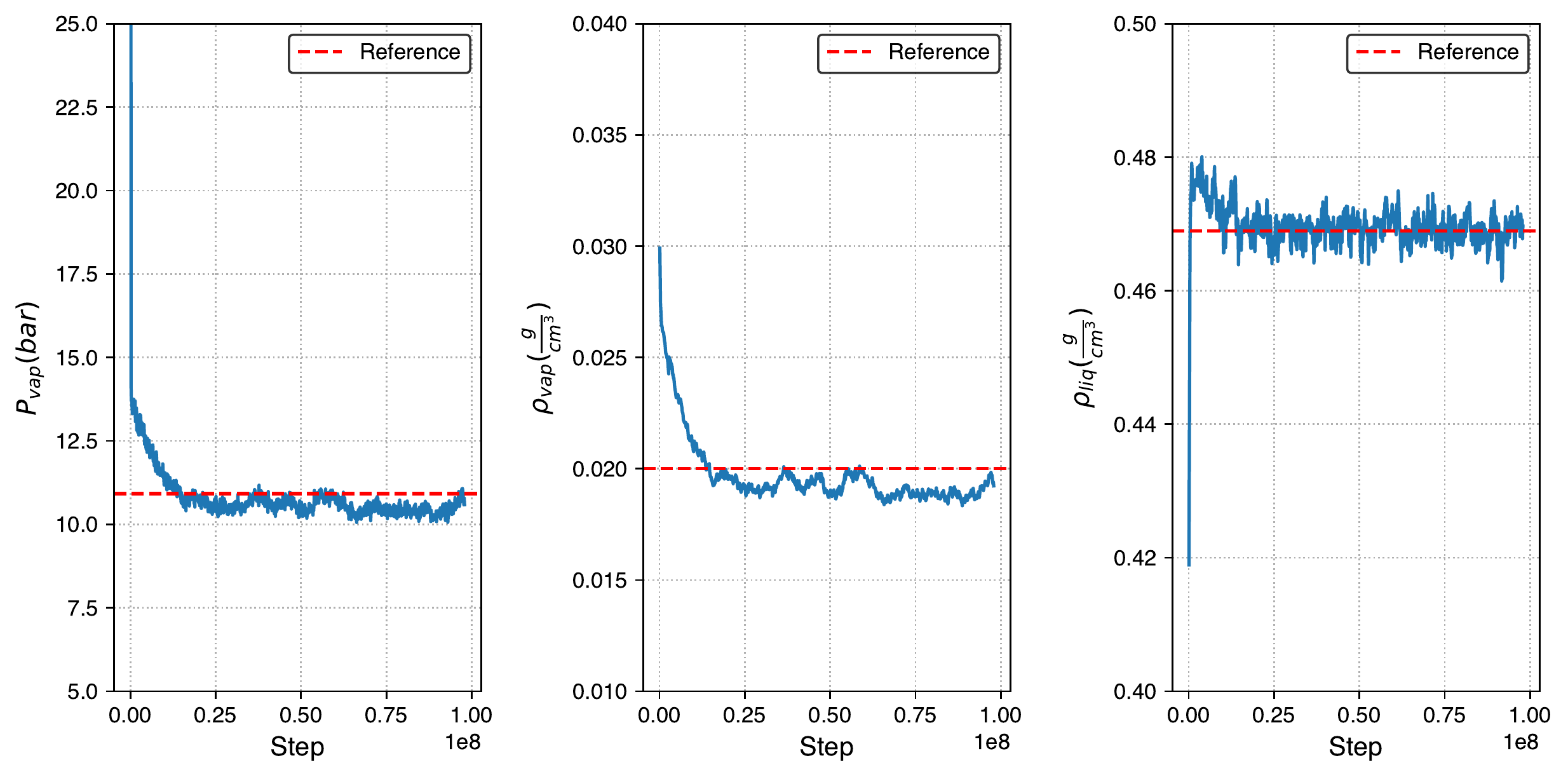}
    \caption{Vapor pressure (left), vapor density (middle), and liquid density
    (right) plots for ethane
    at 236 K, using GEMC in GOMC with the TraPPE force field.}
    \label{fig:vleplots}
\end{figure}

One popular application of molecular simulation involves the use of Monte Carlo (MC) methods, often employing extended ensembles in techniques such as Gibbs Ensemble Monte Carlo (GEMC) or grand canonical Monte Carlo (GCMC), to simulate vapor-liquid equilibria (VLE) properties.
Briefly, GEMC involves simulating two distinct simulation boxes (which generally have different densities and compositions) and performing MC moves to perturb both systems to balance the chemical potentials and pressures between the two simulation boxes\cite{Panagiotopoulos1988,Panagiotopoulos1987}, thus reaching phase equilibrium.
This involves particle displacements within boxes, particle exchanges across boxes, and box volume changes\cite{Panagiotopoulos1988,Panagiotopoulos1987}.
GCMC methods, on the other hand, involve simulating a single simulation box, but performing MC moves to insert or delete particles from a reservoir\cite{Adams1975}.
Additionally, more complex MC moves have been proposed to accelerate equilibration for systems containing complex molecules, including configurational bias Monte Carlo (CBMC) methods\cite{IljaSiepmann1992}.
The \underline{tra}nsferable \underline{p}otentials for \underline{p}hase  \underline{e}quilibria (TraPPE) force field has been designed for conducting simulations for phase equilibria\cite{Martin1998a,Shah2017}.
Here, we present a TRUE workflow that examines ethane vapor-liquid coexistence at a single thermodynamic statepoint.
This workflow utilizes \texttt{mBuild}\cite{Klein2016b,mbuildgithub} to initialize two simulation boxes of ethane (vapor and liquid phases), \texttt{Foyer}\cite{foyergithub,Klein2019} to apply the TraPPE-United Atom (TraPPE-UA) force field\cite{Martin1998a}, and GOMC\cite{GOMChome,GOMC2018,Nejahi2019} to perform a GEMC simulation.
\texttt{mBuild} is used to pack ethane into two different simulation boxes according to the vapor and liquid densities at 236 K (Figure \ref{fig:vleboxes}).
\texttt{Foyer} is used to systematically apply TraPPE force field parameters.
GOMC (version 2.40) is used to perform the GEMC simulation at 236 K, with simulation parameters consistent with the TraPPE implementation\cite{Martin1998a,Shah2017}: Lorentz-Berthelot combining rules, fixed bonds, 1.4 nm Lennard-Jones cutoffs, analytical tail corrections, and Ewald summations for electrostatic interactions.
The resultant analysis validates the vapor pressure, vapor density, and liquid density at 236 K against published reference data (Figure \ref{fig:vleplots})\cite{Martin1998a,Shah2017}.  
A link to this GitHub repository can be found in the supporting information.
All Python dependencies related to building, simulating, and analyzing are openly available and well-documented, and routines are built on top of these dependencies that expose chemistry and statepoint variables.

This workflow is transparent and reproducible, as this workflow and relevant software packages are open-source and available on GitHub\cite{mbuildgithub,foyergithub,GOMC2018,mosdeftrappegithub}.
Furthermore, the workflow is usable by others, as the logged quantities can be analyzed for other properties beyond vapor pressure and densities.
Lastly, this workflow is extensible, as there is a pattern and clear room to implement other state points, molecules, force fields, or simulation engines in addition to implementing workflow managers to facilitate large-scale screening studies.

\subsection{Graphene Slit Pore}\label{graphene}
Graphene has been extensively researched as an electrode material for energy storage applications\cite{Fu2011,Zhan2017,Zhang2019} in recent years mainly due to its high surface area\cite{Meyer2007,Zhan2017,Fu2011}.
Furthermore, the interactions between graphene pores and fluid molecules were studied with MD simulations through the use of slit pore models\cite{Mahurin2016,Feng2011}.
Here we demonstrate a TRUE simulation workflow for a graphene slit pore solvated with aqueous NaCl.
This TRUE graphene simulation was performed with the use of \texttt{mBuild}\cite{Klein2016b,mbuildgithub}, \texttt{Foyer}\cite{Klein2019,foyergithub}, GROMACS\cite{Hess2008,Berendsen1995,Lindahl2001,VanDerSpoel2005,Pronk2013,Abraham2015}, and \texttt{MDTraj}\cite{McGibbon2015MDTraj}.
\texttt{Pore-Builder}\cite{graphenegithub}, an \texttt{mBuild} recipe, was also used to initialize the graphene sheets contained in the system.

\begin{figure}
    \includegraphics[width=\textwidth]{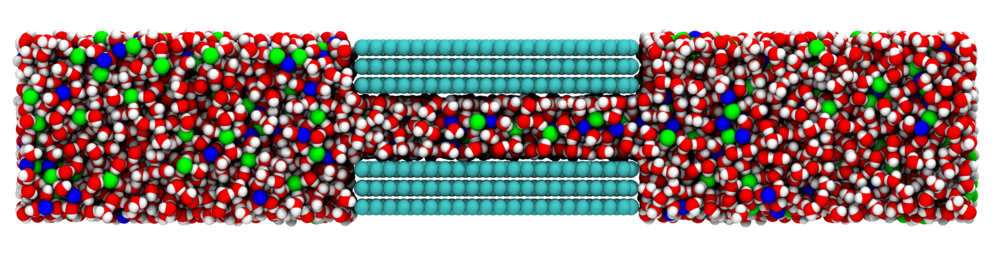}
    \caption{A snapshot of the graphene slit pore system containing graphene
    carbon (cyan), water (red for oxygen and white for hydrogen), sodium ions (blue) and chlorine ions
    (green).}
    \label{fig:graphene_snap}
\end{figure}

This specific system, a graphene slit pore filled with aqueous NaCl, was initialized with \texttt{mBuild}.
\texttt{mBuild} compounds of the specific molecules were initialized with the \texttt{mbuild.load()} function using \texttt{MOL2} files.
Once the molecule compounds were initialized, the \texttt{GraphenePoreSolvent} class within \texttt{Pore-Builder} was utilized.
This specific class makes use of the \texttt{mbuild.Lattice} class and the \texttt{mbuild.solvate} function to build a graphene slit pore system solvated with fluid specified by the user.
In this system, the graphene slit pore was built with three sheets on each side, and a pore width of 1.5 nm.
Additionally, the length of the graphene sheet in the x direction was set to 5 nm and the length of the graphene sheet in the z direction was set to 4 nm.
The bulk region of fluid was set to 6 nm on each side of the slit pore.
5200 waters and 400 Na and Cl ions were solvated into the system.
A snapshot of the system is shown in Figure \ref{fig:graphene_snap}.

Once the graphene slit pore system was initialized as an \texttt{mBuild} compound, it was atom-typed and parametrized with \texttt{Foyer}.
Three force fields were used in this system, with their information stored in three separate \texttt{XML} files: GAFF\cite{Wang2004}, SPC/E\cite{Berendsen1987}, and the force field of Joung and Cheatham (JC)\cite{Joung2008}.
GAFF describes the interactions between the graphene atoms, SPC/E describes the water interactions, and JC describes the Na and Cl interactions.
Each force field uses 12-6 Lennard-Jones interactions, point charges, and harmonic bonds and angles.

The simulation was run with GROMACS 2018.5.
Steepest descent energy minimization was first performed for 1000 steps to remove any energetic clashes from the initial configuration.
Afterwards, a series of two MD simulations were performed with the following parameters: cutoffs of 1.4 nm for Coulombic and van der Waals interactions, a temperature of 300 K controlled with the v-rescale thermostat with a time constant of 0.1 ps, particle mesh Ewald to handle long-range electrostatics, and a timestep of 1 fs.
Additionally, the graphene atoms were frozen in place.
A GROMACS \texttt{NDX} file was created with a \texttt{Water\_and\_ions} group so that the thermostat could be applied to the fluids; no thermostat is applied to the graphene, as the graphene is kept rigid.
First NVT equilibration was performed to further relax the system of any unfavorable configurations for 100,000 steps.
Afterwards, NVT sampling was performed for 2,500,000 steps.
In the sampling run, all bonds were constrained using the LINCS\cite{Hess1997} algorithm.

\begin{figure}
    \includegraphics[width=\textwidth]{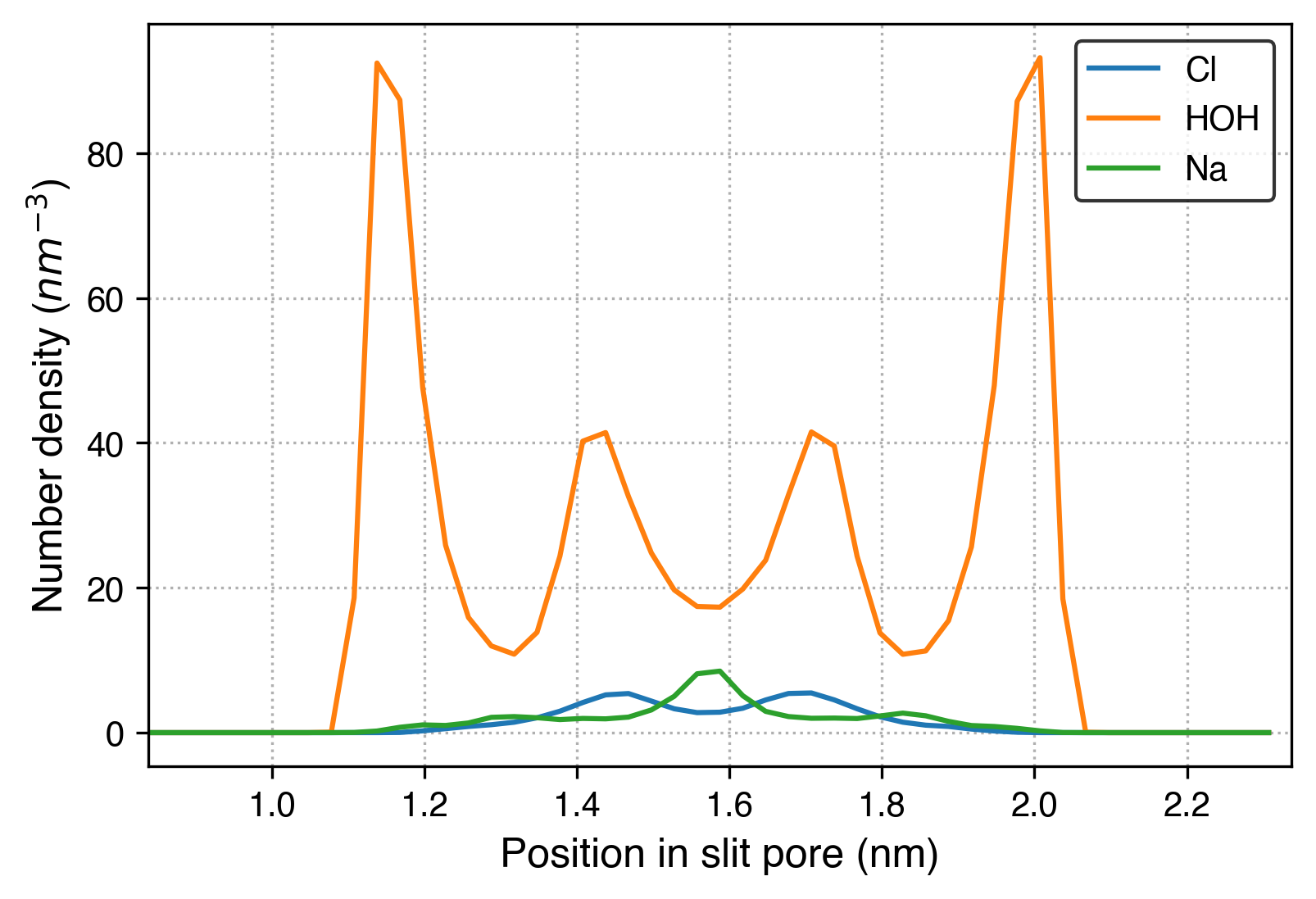}
    \caption{Number Density Profiles across the width of the pore for water, Na,
    and Cl.}
    \label{fig:numberdensity}
\end{figure}

Once the sampling simulation was performed, the number density profile of each fluid type is calculated with the use of \texttt{MDTraj} and plotted with \texttt{Matplotlib}.
The results are shown in Figure \ref{fig:numberdensity}.
From these results, we observe that the water molecules are mainly structured near the pore walls at 1.2 nm and 2.0 nm.
Additionally, there are two smaller peaks around 1.4 and 1.8 nm indicating structuring of water in the middle of the pore.
The Na ions are structured in the middle of the pore around 1.6 nm and the the Cl ions are structured to each side of the Na ions, at around 1.5 and 1.7 nm.
If the graphene was positively or negatively charged, we would expect different structure behavior of the ions.
This simulation can be extended to further understand the effect of various parameters on the fluid structure within the pore.
For example, the user can easily specify a different pore width to study how this impacts the structure of water and ions.
This workflow is encapsulated in a Jupyter notebook, providing the user access to modify any of these high-level parameters.

The workflow for simulating a graphene slit pore satisfies the conditions to be a TRUE simulation.
First, this workflow is transparent as all scripts, input files, and force field information are available for anyone to view\cite{graphene_workflow}.
Next, this workflow is reproducible as the exact steps to set up and run the simulation are contained within a Jupyter notebook.
Barring differences in computer architectures and parallelization schemes, a user running this Jupyter notebook should be able to reproduce the number density profiles from the reference simulation.
Additionally, the trajectories are kept within the workflow directory, allowing users to analyze properties other than number density.
Finally, the functions and classes used to initialize the graphene slit pore system are sufficiently documented so that a user may change characteristics of the simulation if they wish.
For example, a user can extend this workflow to study additional aqueous solutions.

\subsection{Lipid Bilayers}\label{lipids}
MD is a common technique used to perform simulation of biological systems.
An example TRUE biological simulation workflow is demonstrated in the \texttt{true\textunderscore lipids} repository on GitHub\cite{lipids_workflow}.
This workflow focuses on simulating lipids found in the outermost layer of the skin, the stratum corneum (SC).
The SC, which is primarily composed of ceramides (CER), cholesterol (CHOL),  and free fatty acids (FFA) \cite{Weerheim2001}, essentially controls the barrier function of the skin \cite{Madison2003}.
In this workflow a hydrated pre-assembled bilayer of skin lipids configuration was initialized, simulated, and analyzed in a well-documented and reproducible fashion.

\begin{figure}[H]
    \includegraphics[width=\textwidth]{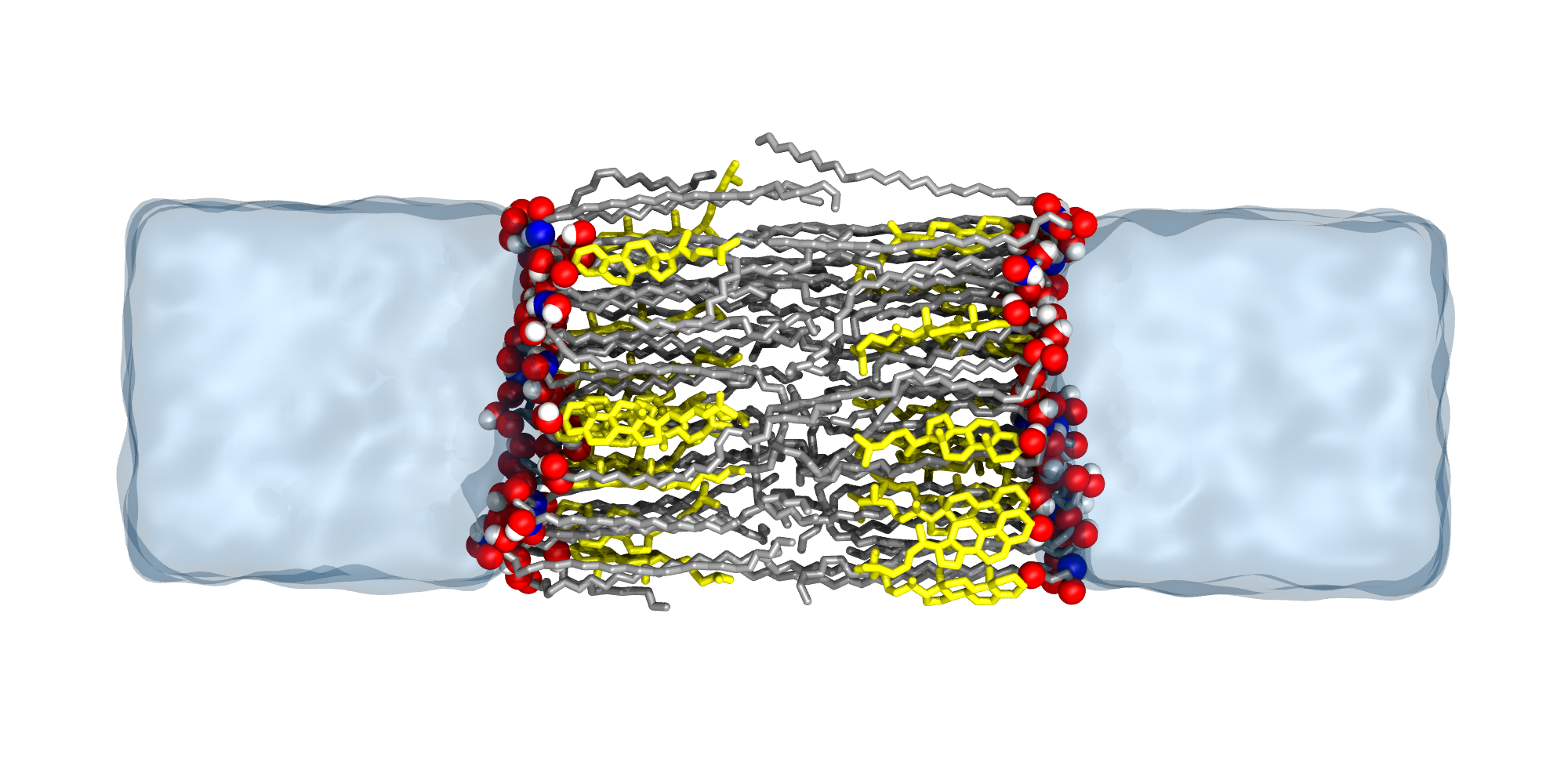}
    \caption{Simulation snapshot of lipid bilayer containing CER N-hydroxysphingosine C24:0 (CER NS), cholesterol and lignoceric acid. The CER NS and FFA tails are shown in silver, cholesterol in yellow, and the headgroup oxygen, nitrogen and hydrogen atoms in red, blue and white respectively.}
    \label{fig:snapshot}
\end{figure}

\texttt{mBuild} was used to initialize the system configuration, specifically utilizing the \texttt{Bilayer}\cite{bilayersgithub} recipe.
A simplified model system containing only CER N-hydroxysphingosine (NS) C24:0, CHOL, and FFA C24:0 was chosen for this example; however a more complex mixture could be easily built by the \texttt{Bilayer} recipe.
For each leaflet of the bilayer, 36 lipids were randomly placed on a 6x6 lattice and rotated about the bilayer normal axis.
The lattice was set up and spaced such that the lateral area occupied by each lipid was equal to the target and as designated by the \texttt{area\textunderscore per\textunderscore lipid} parameter.
In addition, the lipids were rotated about a randomly chosen axis parallel to the bilayer by the tunable \texttt{tilt\textunderscore angle} parameter.
Finally, 20 waters per lipid were added to each of the two ends of the simulation box at a density of 1 $\frac{g}{{cm}^{3}}$. 
The full system contains 72 lipids and 2880 water molecules. 
While many of the steps involved in setting up the initial configuration involve random number generators, exact reproducibility on the same machine was enforced by the initialization of a random seed.

Simulations were conducted using the GROMACS 2018.5 \cite{Hess2008,Berendsen1995,Lindahl2001,VanDerSpoel2005,Pronk2013,Abraham2015} MD engine, using a modified CHARMM36 force field \cite{Guo2013,Klauda2010} with a 1 fs timestep.
The system was first energy-minimized using the steepest descent algorithm for 20000 steps in order to remove high energy atomic contacts.
Temperature fluctuations were stabilized by running a 500 ps NVT simulation using the Berendsen thermostat \cite{BerendsenThermostat} at 305 K with a time constant of 1 ps.
Next, the volume fluctuations were stabilized with a 10 ns NPT simulation at 305 K and 1 atm.
This step and all others hereinafter in this section were in the NPT ensemble and use the Nos\'e-Hoover thermostat \cite{NoseHoover1985} with a time constant of 1 ps and the Parinello-Rahman barostat \cite{ParinelloRahman1981} with a time constant of 10 ps and a compressibility of $4.5\times 10^{-5} \text{bar}^{-1}$.
Still at 1 atm, the system was linearly annealed to 340 K over 5 ns, held at 340 K for 15 ns, linearly cooled to 305 K over 5 ns, and held at 305 K for 25 ns in order to accelerate equilibration of the rotational orientation of the lipids.
Finally, the system was run for 20 ns at 305 K and 1 atm, saving coordinates to a trajectory file every 10 ps.
The final snapshot of the system is shown in figure \ref{fig:snapshot}.
More details on the simulation parameters can be found in the Supporting Information.

\begin{figure}
    \includegraphics[width=\textwidth]{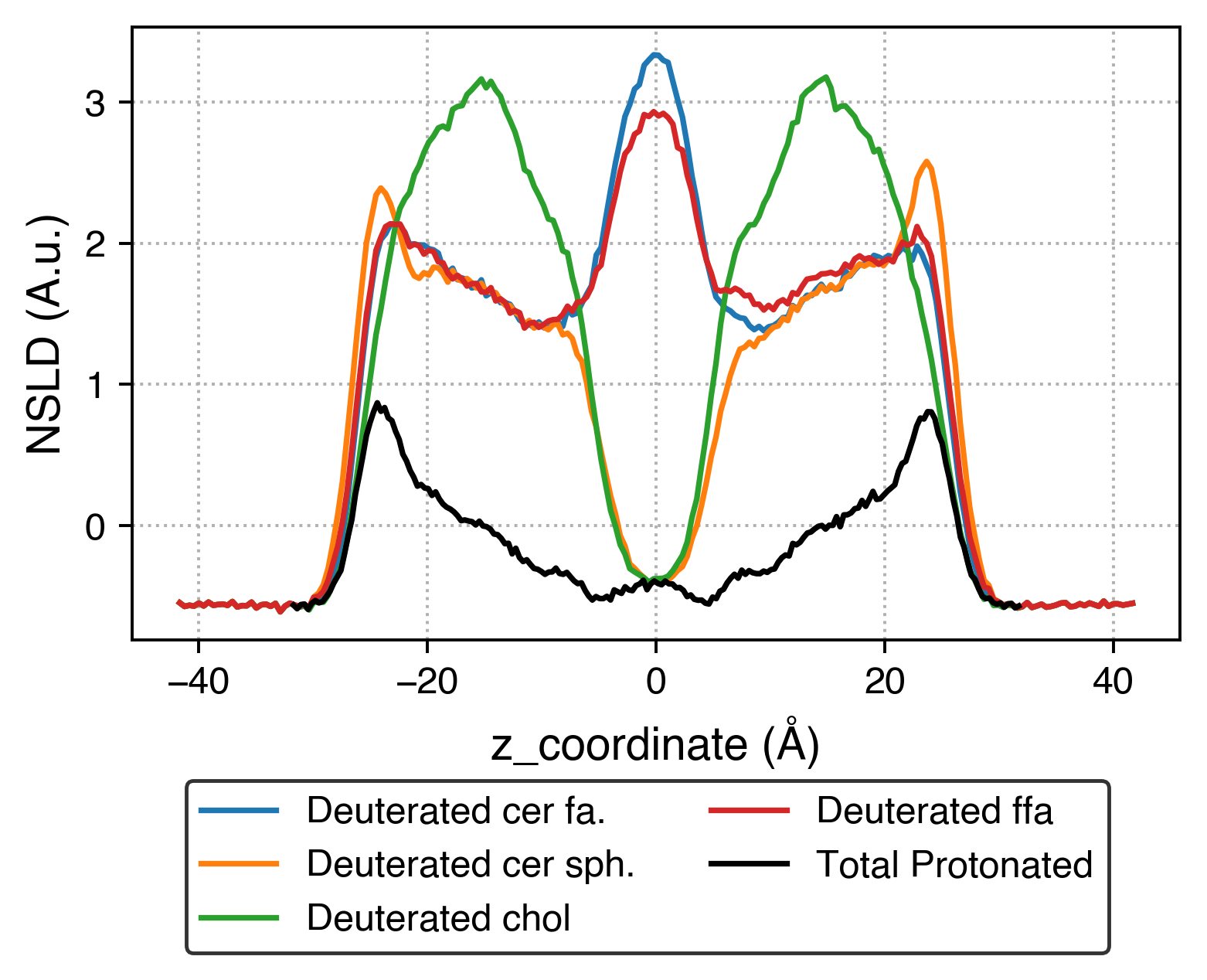}
    \caption{Simulated NSLD profiles for specifically deuterated lipid tails.}
    \label{fig:deuterated}
\end{figure}

The trajectory from the final step was analyzed using \texttt{MDTraj} \cite{McGibbon2015MDTraj}.
Neutron scattering is a popular tool to experimentally obtain structural information of lipid lamella.
A neutron scattering length density (NSLD) profile was calculated for this simulated system along the bilayer normal axis in Figure \ref{fig:deuterated}.

It is apparent from these profiles that the 24-carbon fatty acid tail of the CER and the 24-carbon FFA tail interdigitate, as indicated by the high density peak in the center of the profile.
One can also observe that the 16-carbon sphingosine tail of the CER and CHOL do not interdigitate, and are not present in the middle of the bilayer as there is a low-density trough in their deuteration profiles.
The scattering length densities at the outer edges of the bilayer suggest that the CHOL headgroup is located closer to the center of the bilayer compared to that of other lipids.
In addition to the NSLD profiles, an area per lipid of $32.90 \,\text{\AA}^2$, a tail tilt angle $10.8^\circ$, a nematic order parameter of 0.9414 and a bilayer height of $48.13 \, \text{\AA}$ were calculated in the workflow.

All of these values and plots can be reproduced by executing the workflow.
Furthermore, by extending the workflow to screen over the parameter space, one could identify trends in the calculated values.
The \texttt{Bilayer} recipe is highly modular allowing the user to easily create reproducible bilayer structures containing different lipid types, system sizes, compositions, or water content using an intuitive Python script.

\subsection{Friction Reduction Via Thin Film Coatings}
\begin{figure}[H]
    \includegraphics[width=\textwidth]{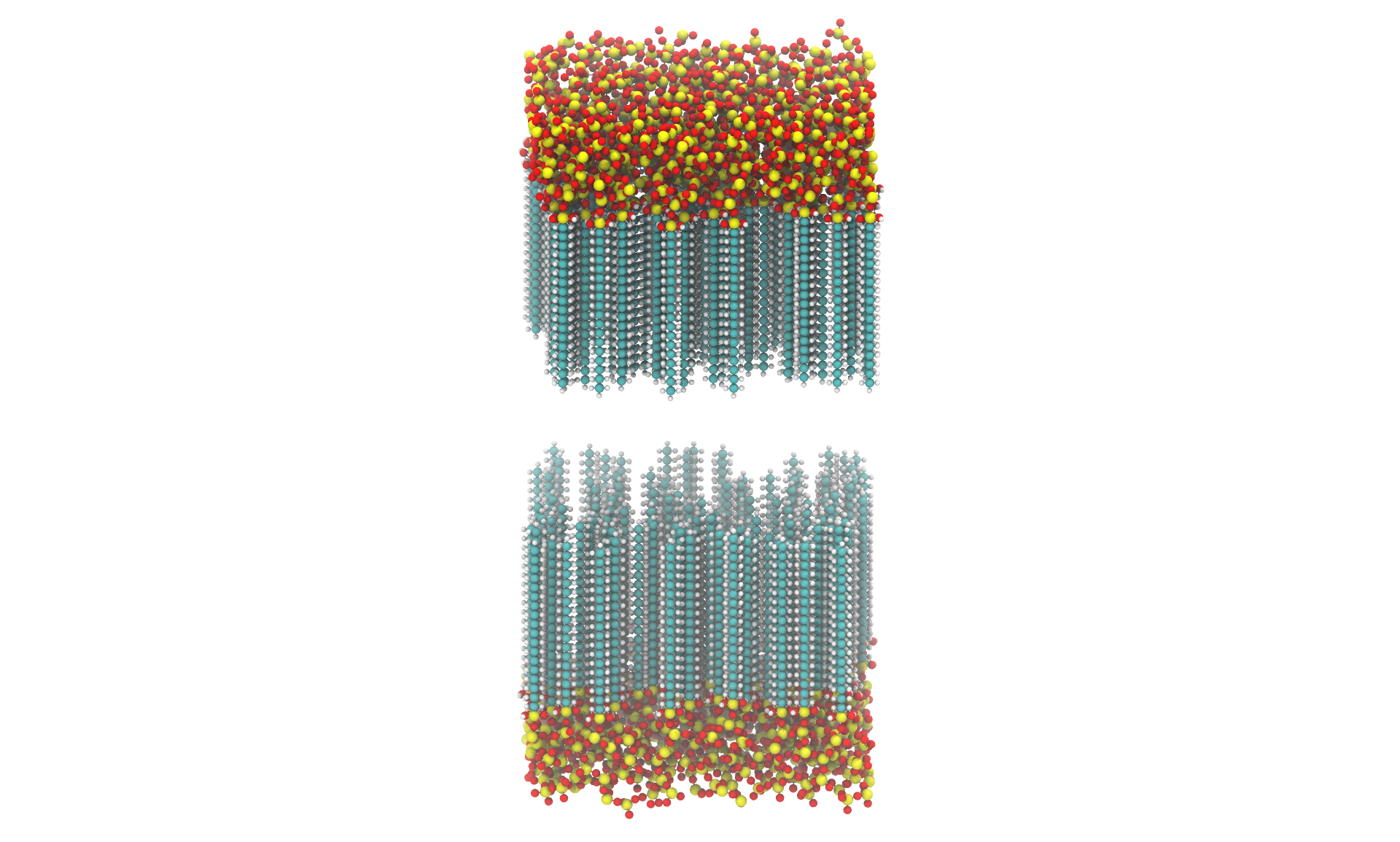}
    \caption{Thin film coated surfaces model}
    \label{fig:Tribology}
\end{figure}

Thin film coatings can be used to modify the surface properties of different systems\cite{Vilt09}.
One potential application of these coatings is to improve tribological properties of surfaces at the micro and nanoscale\cite{Vilt09,Summers16,Summers20}.
In this example, we present a TRUE simulation of a thin film coated system, which is based on an in-depth study by Summers {\em et al.}\cite{Summers20}.
Specifically, we built a system of two $50\times50 \AA$  rectangular silica surfaces, parallel to one another.
Each surface was coated with 100, 17-carbon long, alkylsilane chains, all of which were terminated with a methyl group.
Surface oxygens not functionalized with chains were backfilled with hydrogen caps to form protonated hydrolysis.
These systems were created with \texttt{mBuild}\cite{Klein2016b,mbuildgithub}, and atom-typed, parametrized with \texttt{Foyer}\cite{Klein2019,foyergithub} using OPLS-aa\cite{Jorgensen1996d} force field parameters.
A visualization of the described system is presented in Figure \ref{fig:Tribology}.

\begin{figure}[H]
    \includegraphics[width=\textwidth]{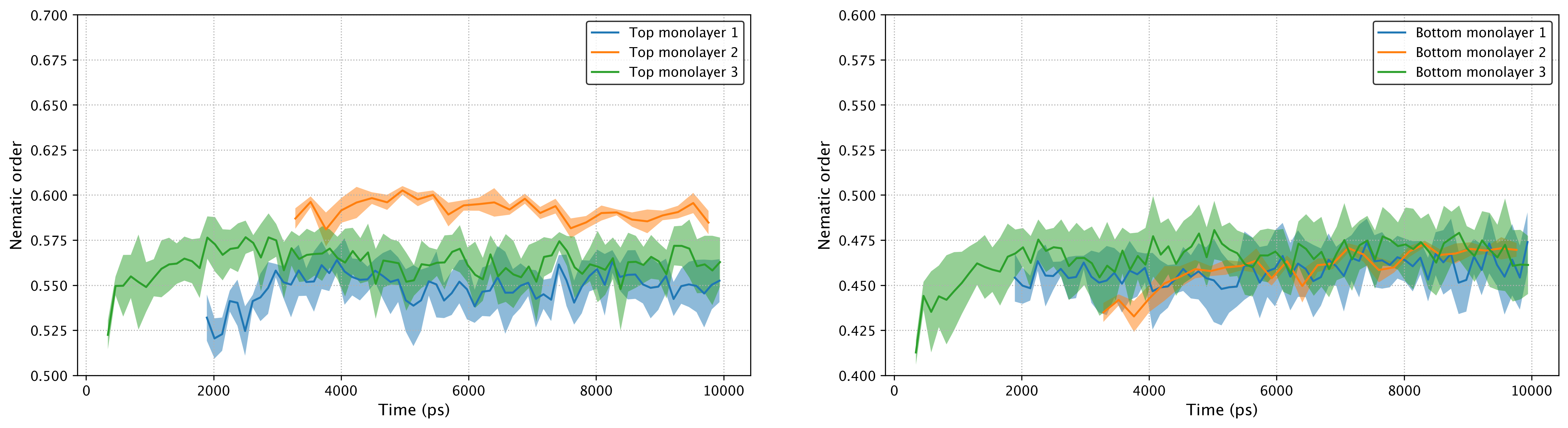}
    \caption{Steady state nematic order of the thin film coated on top and bottom surfaces}
    \label{fig:nematic_order}
\end{figure}

The system was simulated with LAMMPS\cite{PLIMPTON19951} and GROMACS\cite{Hess2008,Berendsen1995,Lindahl2001,VanDerSpoel2005,Pronk2013,Abraham2015}.
MD simulations were run under the canonical ensemble (NVT) and a Nos\'e-Hoover thermostat maintaining the temperature at 298 K\cite{Hoover}.
Long-range electrostatics were calculated using the particle-particle particle-mesh (PPPM) algorithm \cite{Darden}.
The \texttt{rRESPA} time step algorithm was utilized with 0.25 fs, 0.5 fs, 0.5 fs, and 1.0 fs timesteps for bonds, angles, dihedrals, and non-bonded interactions, respectively\cite{Tuckerman}.
The simulation started with energy minimization with LAMMPS for 10,000 steps, followed by another 50,000 steps with GROMACS to bring the monolayers to a more relaxed state.
This process continued with NVT equilibration by GROMACS to bring the system to a desired stable state for 1,000,000 steps.
The workflow proceeded to use GROMACS to compress the system by pulling the top surfaces along the z axis to come into contact with the bottom surface.
In the next step, the top surface was sheared against the bottom surface by imposing a force to pull it along the z axis at the rate of 0.01 $\frac{nm}{ps}$.
The production run was designed to simulate for 5,000,000 steps, which would be equivalent to 10 ns of shearing.
From the GROMACS output file, the properties of the system can be calculated.
The steady-state production period used for final data analysis was determined using the automatic equilibration detection method provided by \texttt{pymbar}\cite{Chodera2016,pymbar}.
By using a defined method to determine the steady-state cutoff, the consistency of the data analysis process can be ensured, holding to the first two criteria of TRUE, transparent and reproducible.
The calculated nematic order of three example runs were determined and are presented in Figure \ref{fig:nematic_order}.

This example focuses on showcasing the extensibility of TRUE, emphasizing the ability to modify and expand the project beyond the original study and parameters of interest.
This goal can be achieved by applying Object-Oriented Programming (OOP) design principles, in combination with open-source libraries.
Encapsulating frequently used code into classes and functions helps improve the reusability of codes and make it easier to create novel systems, just by changing and adding new variables.
By pairing MoSDeF suite libraries, \texttt{mBuild}\cite{Klein2016b,mbuildgithub} and \texttt{Foyer}\cite{Klein2019,foyergithub}, with other open-source libraries, such as \texttt{signac} and \texttt{signac-flow}, part of the \texttt{signac} framework\cite{Adorf2018,Adorf2019}, the extensibility could be made even more manageable and achievable.
Most importantly, all building blocks of the project have to be properly documented, either as comments in the code or in a separate manual.
These practices can help projects expand effectively.
For instance, in this example, although the arguments and variables were defined such that the surfaces were filled with 100, 17-carbon long, alkylsilane chains, each then capped with a methyl group, many unique systems can be  created by altering the chain density, backbone chain length, backbone chemistries, terminal groups, and others as need arises.
The latter part of the example shows how we can expand the project from the original system by varying backbone chain lengths.
For the sake of demonstration, we only show the first few steps of the workflow, starting with setting up the workspace using \texttt{signac}\cite{Adorf2018,Adorf2019}, building corresponding systems with \texttt{mBuild}\cite{Klein2016b,mbuildgithub}, and atom-typing, parametrizing with \texttt{Foyer}\cite{Klein2019,foyergithub}.
Other steps of the simulation can be added analogously.
We implement \texttt{signac-flow}\cite{Adorf2018,Adorf2019} as the workflow manager.
These tools will become vital when needing to run a complete workflow and managing thousands of systems.
All scripts and files needed to run the above example are located in a GitHub repository\cite{tribologygithub}.
Users can interface with this example through the Jupyter notebook located within the repository. 
By providing properly documented codes and scripts used to set up the system, using open-source libraries to perform simulation and data analysis, the first three criteria of TRUE are also satisfied.
This example workflow is an instance of a Transparent, Reproducible, Usable by others, and Extensible, or concisely, TRUE simulation.

\section{Conclusions}\label{sec:conclusions}
In this paper, we have outlined some of the key issues related to reproducibility in molecular simulations of soft matter.
We have also discussed many practices that computational scientists could implement in efforts to result in more reproducible science, such as using scripts instead of manual input, using open-source software tools, and using version control and modern software development practices when developing software.
In this paper, we assert three central claims:

\begin{itemize}
\item The goal in computational molecular science should be simulations that are TRUE: Transparent, Reproducible, Usable by others, and Extensible.
\item Scientific results reported in the literature that depend on molecular simulations should adhere to the above characteristics.
\item Use of the \underline{Mo}lecular  \underline{S}imulation and  \underline{De}sign  \underline{F}ramework (MoSDeF) is one way to enable TRUE simulations.
\end{itemize}

To demonstrate the second claim, we revisit some ``ten rules'' papers\cite{Sandve2013,Elofsson2019,Barba2016} that provide succinct instructions for practicing reproducible science and demonstrate how the above example workflows utilize MoSDeF to this end.
A common recommendation in these discussions is that every step in a workflow should be automated and free of manual input, i.e. scriptable.
MoSDeF, in its current state, is a set of Python libraries designed specifically to address this.
In a single Python script (or Jupyter notebook), each step of a molecular simulation workflow (generation of particle coordinates, application of a force field, running of a molecular simulation, and analysis of the results) can be specified and run.
The objective of measuring physical properties from some chemical input can be achieved with one call to an executable (although the simulation may take some hours or days to run).
In order for these scripts, which include many imports to other Python libraries, to produce identical (or sufficiently identical) results some years in the future, the underlying libraries must be version-controlled.
The core MoSDeF packages (\texttt{mBuild} and \texttt{Foyer}) undergo regular releases, tagged with semantic version numbers, every few weeks or months as they are developed.
Other packages, such as simulation engines, the packages in the signac framework, and underlying scientific Python packages, are also version-controlled and undergo regular releases.
Specifying the version of each software package used in a simulation workflow is not necessarily sufficient to ensure reproducible science, but it is a significant improvement over the use of \textit{ad hoc} or in-house scripts that often lack version control, proper testing, or releases.
Similarly, it has been argued that the use of community-developed software libraries, and the extension of such libraries, further promotes reproducibility as compared to closed-source, in-house development\cite{Donoho2009}.
MoSDeF is a set of open-source that interface with other open-source, community-developed libraries and software tools.

Additionally, MoSDeF makes use of virtually no GUIs - or, more specifically, no GUIs that hide the details of a simulations protocol from the user.
Some molecular visualization tools (\texttt{NGLview}, \texttt{py3DMol}, \texttt{VMD}, \texttt{ovito}, \texttt{fresnel}) can be used in conjunction with MoSDeF, but these are only tools to visualize systems and do not hide workflow details or replace steps in a workflow.

Finally, we would like to discuss an additional benefit of shifting toward more reproducible computational studies: the facilitation of large-scale screening of physiochemical space.
Continuous improvements in computer hardware and recent advancements in machine learning methodologies have driven interest in studying large data sets, typically many orders of magnitude larger than typically seen in the literature.
Provided that each step in a workflow can be automated - in other words, scriptable with no manual input - a single simulation can be repeated with different physical inputs (e.g. at different thermodynamic statepoints or with different chemistries) by only modifying the input parameters.
For example, consider some system at temperature and pressure $(T, P)$ for which we care about some physical property $A$.
One can run a simulation at $(T_1, P_1)$ and get property $A_1$ but later decide we want to look at some other temperature and/or pressure.
One could manually move some files around and get property $A_2$ from statepoint $(T_2, P_2)$ without prohibitive trouble, but doing this once is a plausible source of human error and repeating this process many times is not feasible.
Screening over $N$ statepoints in a reproducible manner necessitates that running a workflow at a single statepoint is reproducible.
We hope the practices outlined in this paper and the use of MoSDeF can enable reproducible computational science at each scale.

\section{Acknowledgments}
The development of MoSDef, as reported in this paper, was primarily supported by the National Science Foundation (NSF)) through grant OAC-1835874 ``Software for Building a Community-Based Molecular Modeling Capability Around the Molecular Simulation Design Framework (MoSDeF).'' 
Earlier developments leading to MoSDeF were supported by previous NSF grants CBET-1028374 "Cyber-Enabled Design of Functional Nanomaterials", OAC-1047828 ``Development of an Integrated Molecular Design Environment for Lubrication Systems (iMoDELS)'' and OAC-1535150 ``Development of a Software Framework for Formalizing Forcefield Atom-Typing for Molecular Simulation.''
The development of code within MoSDeF specifically relevant to energy storage systems and the example described in Section~\ref{graphene} were supported as part of the Fluid Interface Reactions, Structures and Transport (FIRST) Center, an Energy Frontier Research Center funded by the U.S. Department of Energy, Office of Science, Office of Basic Energy Sciences. 
The development of code within MoSDeF specifically relevant to self-assembling lipid systems and the example described in Section~\ref{lipids} were supported by National Institute of Arthritis and Musculoskeletal and Skin Diseases grant R01AR072679 ``Insights into Skin Barrier Function: In Silico and Experimental Studies of Healthy and Diseased Stratum Corneum Lipid Models.''
We also acknowledge key contributions to MoSDeF by former group members Christoph Klein (myriad.com) and Andrew Z. Summers (enthought.com) and by ISIS staff Peter Volgesi, Umesh Timalsina, and Janos Sallai.

\section{Supplemental Information}\label{sec:si}
\subsection{Packages and Libraries Necessary to Run Example TRUE Simulations}
To successfully run these examples in a TRUE fashion requires the methodology to be reproducible as well as the software used for \emph{all} steps of the example/study.
Without this, changes in various software packages and their dependencies can introduce another source of irreproducibility to a TRUE study.
Contained below is a detailed listing of the main software packages and libraries used throughout the examples.
This suite of software is intended to be installed partly with the \texttt{conda} scientific software package manager, other python modules not accessible through \texttt{conda} will be installed from their source code, and finally, any simulation engine/extraneous packages will be compiled from their source code as well.
Comprehensive installation instructions will be provided for each step of this process and annotated.
Due to limited molecular simulation engines and other libraries being accessible with the \texttt{Windows} operating system, these next steps are only expected to run successfully on \texttt{GNU/Linux} and Apple \texttt{MacOS} operating systems.

The following text assumes the reader intends to install these packages on their local machine or compute node and can access a terminal emulator.

\subsection{Installation of the \texttt{conda} Package Manager}

To install the \texttt{conda} package manager, run the following commands in your shell session if you are using a MacOS operating system.

\emph{The \texttt{\$} denotes a line in your terminal emulator and is not part of the command.}
\begin{lstlisting}
  $ cd ${HOME}
  $ curl -O https://repo.anaconda.com/miniconda/Miniconda3-latest-MacOSX-x86_64.sh
  $ /bin/bash Miniconda3-latest-MacOSX-x86_64.sh
\end{lstlisting}

If you are using a local GNU/Linux machine, the following commands should be executed.
\begin{lstlisting}
  $ cd ${HOME}
  $ curl -O https://repo.anaconda.com/miniconda/Miniconda3-latest-Linux-x86_64.sh
  $ /bin/bash Miniconda3-latest-Linux-x86_64.sh
\end{lstlisting}

Follow the prompts according to your installation preferences, although the default location (your home directory) can be expected to work.
Please refer to the documentation (\url{https://conda.io/projects/conda/en/latest/user-guide/index.html}) for any additional help or if your installation is on a computing cluster.

\subsection{Creating the \texttt{conda} Environment}
After following the previous instructions and initalized the \texttt{conda init} shell support (if you are unsure what that is, refer to the user guide above).
Now we will install the software and libraries needed to follow along with the TRUE examples.

Begin by creating a new \texttt{conda} environment with \texttt{Python 3.7} as the base \texttt{Python} interpreter, and activating into that environment.
A \texttt{conda} environment is a directory that contains all of the necessary libraries and software needed based on your installation instructions.
For example, you can have multiple environments, all of which use a different version of \texttt{Python}, and \texttt{conda} allows you to swap between these environments by \emph{activating} (switching to) or \emph{deactivating} (exiting) an environment.
These environments are independent of one another, so modifying a certain environment will not change any other environments' installed packages.
All of the examples in this paper are expected to be ran while you are in the \texttt{true37} python environment.
\begin{lstlisting}
  $ conda create -n true37 python=3.7
  $ conda activate true37
\end{lstlisting}

Next, we will install all of our conda-installable packages and dependencies.
\begin{lstlisting}
  $ conda install -c conda-forge -c mosdef -c omnia -c bioconda mbuild foyer signac signac-flow hoomd gromacs=2018.4 lammps pandas matplotlib unyt py3dmol scipy openbabel gsd
\end{lstlisting}

An alternate option is to add the channels that conda will search to resolve the installation procedure to the \texttt{.condarc} file located in your home directory.
\begin{lstlisting}
  $ conda config --add channels conda-forge
  $ conda config --add channels mosdef
  $ conda config --add channels omnia
  $ conda config --add channels bioconda
  $ conda install mbuild foyer signac signac-flow hoomd gromacs=2018.4 lammps pandas matplotlib unyt py3dmol scipy openbabel gsd
\end{lstlisting}

To list all of the installed packages in your current \texttt{conda} environment, run: 
\begin{lstlisting}
  $ conda list
\end{lstlisting}

\subsection{Create a Temporary Workspace}
Note that we are also creating a master directory where all of these TRUE examples will be stored, do not run a \texttt{git clone} command while inside another \texttt{git} repository.
The commands to make the master directory and changing to that directory are idempotent, so you can copy and paste the 3 commands below as many times as desired.

\begin{lstlisting}
  $ export TRUE_EXAMPLES=${HOME}/true_examples
  $ mkdir -p ${TRUE_EXAMPLES}
  $ cd ${TRUE_EXAMPLES}
\end{lstlisting}

\subsection{Installation of the \texttt{mosdef\_trappe} TRUE example}
This example makes use of the Monte Carlo engine \texttt{GOMC} (\url{https://github.com/GOMC-WSU/GOMC.git}).

\texttt{GOMC} also requires a working c/c++ compiler, please consult the user manual: (\url{https://gomc-wsu.github.io/Manual/software_requirements.html}) for additional help.

The instructions below assume you have an accessible c++ compiler.

The next step is to download a version of this sample workflow and install any dependencies, and to do that we will use the \texttt{git} version control tool.
MacOS and GNU/Linux ship with a version of \texttt{git}, the commands to run are listed below.

To begin, we must compile and install \texttt{GOMC}.

\begin{lstlisting}
  $ export TRUE_EXAMPLES=${HOME}/true_examples
  $ mkdir -p ${TRUE_EXAMPLES}
  $ cd ${TRUE_EXAMPLES}

  # GOMC requires cmake for compilation, we will install it from conda
  $ conda activate true37
  $ conda install -c conda-forge cmake

  # clone GOMC
  $ git clone https://github.com/GOMC-WSU/GOMC.git
  $ cd GOMC
  $ chmod u+x metamake.sh
  $ ./metamake.sh
  # once compiled, the executable should be located in the bin directory
  $ ls ./bin

  # add the gomc bin folder to our path, so we can find the executable no matter the directory
  $ LOC_GOMC="$(pwd)/bin"
  $ export PATH="${LOC_GOMC}:$PATH"
\end{lstlisting}

After installing \texttt{GOMC}, we can finally install \texttt{mosdef\_trappe}.
\begin{lstlisting}
  $ export TRUE_EXAMPLES=${HOME}/true_examples
  $ mkdir -p ${TRUE_EXAMPLES}
  $ cd ${TRUE_EXAMPLES}
  $ LOC_GOMC="${TRUE_EXAMPLES}/GOMC/bin"
  $ export PATH="${LOC_GOMC}:$PATH"

  $ git clone https://github.com/ahy3nz/mosdef_trappe.git
  $ cd mosdef_trappe
  $ conda activate true37
  $ python -m pip install -e .
\end{lstlisting}

\subsection{Installation of the \texttt{true\_graphene} Example}
This TRUE example requires a few dependencies, including a \texttt{mbuild} plugin as well

\begin{lstlisting}
  $ export TRUE_EXAMPLES=${HOME}/true_examples
  $ mkdir -p ${TRUE_EXAMPLES}
  $ cd ${TRUE_EXAMPLES}

  # install the mbuild plugin
  $ git clone https://github.com/rmatsum836/Pore-Builder.git
  $ cd Pore-Builder
  $ conda activate true37
  $ conda install -c conda-forge -c mosdef -c omnia --file ./requirements.txt
  $ python -m pip install -e .
  # now clone the graphene_pore example
  $ git clone https://github.com/rmatsum836/true_graphene.git
\end{lstlisting}

\subsection{Installation of the \texttt{true-lipids} example}
By installing the other packages above, all of the dependencies for this example should be installed.

\begin{lstlisting}
  $ export TRUE_EXAMPLES=${HOME}/true_examples
  $ mkdir -p ${TRUE_EXAMPLES}
  $ cd ${TRUE_EXAMPLES}

  $ conda activate true37
  $ git clone https://github.com/uppittu11/true_lipids.git
\end{lstlisting}
If running this example does not work, follow this installation step below.

\begin{lstlisting}
  $ export TRUE_EXAMPLES=${HOME}/true_examples
  $ mkdir -p ${TRUE_EXAMPLES}
  $ cd ${TRUE_EXAMPLES}

  $ conda activate true37
  $ cd true_lipids
  $ conda install -c conda-forge -c mosdef -c omnia mbuild mdtraj py3dmol
\end{lstlisting}

\subsection{Installation of the TRUE-nanotribology Project}
To access and install the necessary software for this repository, the following steps should be taken.

\begin{lstlisting}
  $ export TRUE_EXAMPLES=${HOME}/true_examples
  $ mkdir -p ${TRUE_EXAMPLES}
  $ cd ${TRUE_EXAMPLES}

  $ conda activate true37
  $ git clone https://github.com/daico007/TRUE-nanotribology.git
  $ cd TRUE-nanotribology
  $ conda install -c conda-forge -c mosdef -c omnia -c bioconda --file ./requirements.txt
\end{lstlisting}

\subsection{Removing the Installed Software}

Listed below are all the steps needed to remove the examples, as well as the conda environment that was created.
Refer to \texttt{miniconda}'s site for assistance unsintalling miniconda.

\begin{lstlisting}
  $ export TRUE_EXAMPLES=${HOME}/true_examples
  # The rm -f (force) command might be needed to remove the directories
  $ rm -r ${TRUE_EXAMPLES}

  # remove the conda environment, and clean up
  $ conda activate base
  $ conda remove -n true37 --all

  $ conda clean --index-cache --packages --tarballs --yes

\end{lstlisting}

\end{document}